\documentclass[12pt,intlim,righttag]{article}

\usepackage{amscd,amsfonts,amssymb,amsmath,,amsthm,latexsym,array,hhline}

\newcommand\beq{\begin{equation}}
\newcommand\eeq{\end{equation}}

\newcommand{\sg}{\sigma}
\newcommand{\ld}{\lambda}

\newcommand{\td}{\tilde}

\newcommand{\kap}{\kappa}

\newcommand{\be}{\begin}
\newcommand{\ee}{\end}
\newcommand{\lbl}{\label}
\newcommand{\ds}{\displaystyle}

\theoremstyle{Theorem}

\newtheorem{prop}{\indent Proposition}[section]

\theoremstyle{definition}

\theoremstyle{remark}

\begin{document}

\title{Robust utility maximization for diffusion market model with misspecified coefficients}

\author{R. Tevzadze $^{1),3)}$ and T. Toronjadze $^{1),2)}$}
\date{~}
\maketitle

\begin{center}
$^{1)}$ Georgian--American University, Business School, 3,
Alleyway II,\\ Chavchavadze Ave. 17, a, Tbilisi, Georgia,

E-mail: toronj333@yahoo.com  \\[2mm]
$^{2)}$ A. Razmadze Mathematical Institute, 1, M. Aleksidze St.,
Tbilisi, Georgia,  \\[2mm]
$^{3)}$Institute of Cybernetics, 5, S. Euli St., Tbilisi, Georgia,\\
E-mail: tevzadze@cybernet.ge
\end{center}

\numberwithin{equation}{section}
\begin{abstract}

The paper studies the robust maximization of utility of terminal
wealth in the diffusion financial market model.  The underlying
model consists  with risky tradable asset, whose price is described by diffusion process with misspecified trend and
volatility coefficients, and non-tradable asset with a
known parameter. The robust utility functional is defined in
terms of a HARA utility function. We give explicit
characterization of the solution of the problem by means of a
solution of the HJBI equation.

\bigskip

\noindent {\bf Key words and phrases}:{The maximin problem, saddle
point, Hamilton-Jacobi-Bellman-Isaacs (HJBI) equation, robust
utility maximization, generalized control.}

\noindent {\bf  Mathematics Subject Classification (2000)}: 60H10,
60H30, 90C47.
\end{abstract}
\

\section{Introduction}

 The purpose of the present paper is to study the robust maximization of
utility of terminal wealth in the diffusion financial market model, where
appreciate rate and volatility of the asset price are not known
exactly.

The utility maximization problem was first studied by Merton
(1971) in a classical Black-Scholes model. Using the Markov
structure of the model he derived the Bellman equation for the
value function of the problem and produced the closed-form
solution of this equation in cases of power, logarithmic and
exponential utility functions.

For general complete market models, it was shown by Pliska (1986),
Cox and Huang (1989) and Karatzas et al (1987) that the optimal
portfolio of the utility maximization problem is (up to a
constant) equal to the density of the martingale measure, which is
unique for complete markets. As shown by He and Pearson (1991) and
Karatzas et al (1991), for incomplete markets described by
Ito-processes, this method gives a duality characterization of
optimal portfolios provided by the set of martingale measures.
Their idea was to solve the dual problem of finding the suitable
optimal martingale measure and then to express the solution of the
primal problem by convex duality. Extending the domain of the dual
problem the approach has been generalized to semimartingale models
and under weaker conditions on the utility functions by Kramkov
and Schachermayer (1999). All these papers consider a model of
utility which assumes that beliefs are represented by a
probability measure. In 1999, Chen and Epstein have introduced a
continuous-time intertemporal version of multiple-priors utility
in the case of a Brownian filtration. In this case, beliefs are
represented by a set $\cal P$ of probability measures and the
utility  is defined as the minimum of expected utilities over the
set $\cal P$. Independently, Cvitanic (2000) and Cvitanic and
Karatzas (1999) examine, for a given option, hedging strategies
that minimize the expected ``shortfall'' that is, the difference
between the payoff and the terminal wealth. They consider the
problem of determining a ``worst-case'' model $\td Q$, that is a
model that maximizes the minimal shortfall risk over all possible
priors $Q \in{\cal P}$. They show that under some assumptions
their sup-inf problem can be written as an inf-sup problem. In
2004 Quenez studied the problem of utility maximization  in an
incomplete multiple-priors model, where asset prices are
semimartingales. This problem corresponds to a sup-inf problem
where the supremum is taken over the set of feasible wealths $X$
(or portfolios) and where the infimum is taken over the set of
priors~${\cal P}$. The author showed that, under suitable
conditions, there exists a saddle-point for this problem.
Moreover, Quenez developed a dual approach which consists in
solving a dual minimization problem over the set of priors and
supermartingale measures and showed how a solution of the dual
problem induces one for the primal problem.

These sup-inf problems also can be called robust optimization
problems since optimization involves an entire class $\cal P$ of
possible probabilistic models and thus takes into account model
risk. Optimal investment problems for such robust utility
functionals were considered, among others, by Talay and Zheng
(2002), Korn and Wilmott (2002), Quenez (2004), Schied
(2005),(2008),  Korn and Menkens (2005), Gundel (2005),Bordigoni
at al. (2007),  Schied and Wu (2005), F\"ollmer and Gundel (2006),
Hern\'andez-Hern$\acute{\rm a}$ndez and Schied (2006, 2007) .

The numerous of publications are concerned to the case when one of
these parameters is known exactly. In the case of unknown drift
coefficient the existence of saddle point of corresponding minimax
problem has been established and characterization of the optimal
strategy has been obtained (see \cite{CvKa}, \cite{herher},
\cite{gun}). For the case of unknown volatility coefficients the
construction of hedging strategy  were given in the works
\cite{ave}, \cite{ah}, \cite{ahn}, \cite{lato}.

The most difficult case is to characterize the optimal strategy of
minimax (or maximin) problem under uncertainty of both drift and
volatility terms.
 Talay and Zheng \cite{TaZ} applied the PDE-based approach to
 the max-min problem  and characterized
 the value as a viscosity solution of corresponding
Hamilton-Jacob-Bellman-Isaacs (HJBI) equation. In general such
problem does not admit a
 saddle point.

We consider incomplete diffusion financial market model which
resembles to the model considered by  Schied (2008),
Hern\'andez-Hern$\acute{\rm a}$ndez and Schied (2006, 2007). We
suppose that the market consists with riskless asset, risky
tradable asset whose trend and volatility are misspecified and
non-tradable asset with a known parameters. Different from Quenez
(2004) and Schied (2008) approach  we solve the maximin problem
using HJBI equation which corresponds to the primal problem. In
case of unknown trend and volatility coefficient such maximin
problem doesn't have a saddle point in general. We are extending
the set of model coefficients i.e. doing some ``randomization''
and by this way we are getting a problem with the saddle point.
This gives us the possibility to replace maximin problem by
minimax problem which is convenient to study HJBI equation
properties. Particularly, we have found such form of this equation
which coincides to equation obtained by
Hern\'andez-Hern$\acute{\rm a}$ndez and Schied (2006) in case of
known volatility. The solvability in classical sense of obtained
equation is established and in case of specific drift coefficient
HJBI equation is explicitly solved and saddle point (optimal
portfolio and optimal coefficients) of maximin problem found as
well.

The paper is organized as follows. In section 2, we describe the
model and consider the misspecified coefficients as a generalized
controls. Further we show the existence of saddle point of
generalized max-min problem and derive HJBI equation for value
function. In section 3 we prove the solvability in the classical
sense of obtained PDE in the case of power utility and give
explicit PDE-characterization of a robust maximization problem.

\section{The generalized coefficients and existence of saddle point}

Suppose that the financial market consists in a riskless asset
\beq dS_t^0=r(Y_t)S_t^0dt\eeq with $r(y)\ge0$ and risky financial
assets whose prices defined through stochastic differential
equation (SDE)
\begin{align}\label{mar}
\frac{dS_t}{S_t}=(\td b(Y_t)+\mu_t)dt+\sigma_tdw_t.
\end{align}
Here $w_t$ is a standard Brownian motion and $Y_t$ denotes a
return of non-traded asset modelled by SDE
\begin{align}\label{mar1}
dY_t=\beta(Y_t)dt+\left(\rho dw_t+\sqrt{1-\rho^2}dw_t^\bot\right),
\end{align}
for a some correlation factor $\rho\in[0,1]$ and standard Brownian
motion $w^\bot$, which is  independent of $w$. We note $b=\td b-r$
and assume that

\smallskip
A1) $b(y),\;\beta(y),\;r(y)$ belongs to $C_b^1(\mathbb{R})$,

\smallskip
A2) $b^\prime(y)$ belongs to $C_0(\mathbb{R})$,

\smallskip
\noindent where $C_b^1(\mathbb{R})$ denotes the class of
 bounded
continuous functions with bounded derivatives and
$C_0(\mathbb{R})$ denotes the class of continuous function with
compact supports.

 Introduce the set  $\tilde {\cal U}_K$ of all measurable process
 $(\mu_t,\sg_t)$ with value in the set $K=[\mu_-,\mu_+]\times[\sg_-,\sg_+],$
 where $0\le\mu_-\le\mu_+,\;0\le\sg_-\le\sg_+$
 and  denote by ${\cal U}_K$ the subset of
 predictable processes from $\tilde {\cal U}_K$. By $\Pi_x$ we denote the set of predictable processes
 such that
 $\int_0^T\pi_t^2dt<\infty,\;P-a.s.$ and corresponding wealth process , defined as a solution of
 SDE
\begin{align}
dX_t&=(1-\pi_t)X_t\frac{dS_t^0}{S_t^0}+\pi_tX_t\frac{dS_t}{S_t},\\
X_0&=x,
\end{align}
satisfies condition $X_t(\pi)\ge 0$.

 The objective of economic agent is to find the optimal robust strategy
of the problem
\begin{align}\label{mima}
\max_{\pi\in\Pi_x}\min_{(\mu,\sigma)\in{\cal
U}_K}EU(X_T^{\mu,\sigma}(\pi)),
\end{align}
with
\begin{equation} \label{cap1}
\begin{aligned}
dX_t& =r(Y_t)X_tdt+\pi_t(b(Y_t)+\mu_t)dt+\pi_t\sigma_tdw_t,\;\; X_0=x,\\
dY_t & =\beta(Y_t)dt+ \rho dw_t+\sqrt{1-\rho^2}dw_t^\bot,\;\;
\eta_0=y,
\end{aligned}
\end{equation}
where $U(x)$ is HARA\footnote{$^)$ The function $U(\cdot)$ is a HARA (Hyperbolic Absolute Risk Aversion)
utility if $-u_2'(x)/u'(x)=\gamma/x$, $\gamma<1$, $\gamma\neq 0$}${}^)$ utility function.

If we denote by $\nu_t(d\mu d\sg)$ the regular conditional
distribution of the pair of processes $(\mu,\sg)\in \tilde {\cal
U}_K$, with respect to filtration ${\cal F}_t$ and by $(f,\nu_t)$
the integral $\int_Kf(\mu,\sg)\nu_t(d\mu d\sg)$, where
$f(\mu,\sg)$ is an arbitrary continuous function , we can perform
the following extension maximin problem
\begin{gather}\label{cap22}
\max_{\pi\in\Pi_x}\min_{(\mu,\sigma)\in\td{\cal U}_K}EU(X_T^{\mu,\sigma}(\pi)),\\
\begin{aligned}
dX_t& =r(Y_t)X_tdt+\pi_t(b(Y_t)+(\mu,\nu_t))dt+\pi_t\sqrt{(\sigma^2,\nu_t)}dw_t,\;\; X_0=x,\\
dY_t& =\beta(Y_t)dt+
\rho\frac{(\sigma,\nu_t)}{\sqrt{(\sigma^2,\nu_t)}}dw_t+\sqrt{1-\rho^2\frac{(\sigma,\nu_t)^2}{(\sigma^2,\nu_t)}}dw_t^\bot,\;\;
\eta_0=y.
\end{aligned}
\end{gather}
Introduce   the set ${\cal P}(K)$ of probability distributions with support on $K$ (${\cal P}(K)$ is a compact metric space in a week topology,
see \cite{part}).
Denote by $\nu_t$ the ${\cal P}(K)$-valued  predictable process. Such type process usually called the generalized control in control theory.
From now on we identify $\td {\cal U}_K$ to the set of
generalized controls.

\be{rem} Let ${}^pY$ be the predictable projection of a process
$Y$ (see \cite{L-Sh2}). Then for  $(\mu_t,\sg_t),\;(\mu,\sg)\in
\tilde {\cal U}_K$ we have the equalities
${}^p\mu_t=(\mu,\nu_t),\;{}^p\sg_t=(\sg,\nu_t)$ and we can write
\begin{equation}\label{cap2}
\begin{aligned}
dX_t& =r(Y_t)X_tdt+\pi_t(b(Y_t)+{}^p\mu_t)dt+\pi_t\sqrt{{}^p\sigma_t^2}dw_t,\;\; X_0=x\\
dY_t& =\beta(Y_t)dt+
\rho\frac{{}^p\sg_t}{\sqrt{{}^p\sg_t^2}}dw_t+\sqrt{1-\rho^2\frac{\left({}^p\sg_t\right)^2}{{}^p\sg_t^2}}dw_t^\bot,\;\;
\eta_0=y.
\end{aligned}
\end{equation}
\ee{rem} \qed

Since
\begin{align}
\notag\be{pmatrix}
\pi_t\sqrt{(\sigma^2,\nu_t)}&0\\[3mm]
\rho\frac{(\sigma,\nu_t)}{\sqrt{(\sigma^2,\nu_t)}}&\sqrt{1-\rho^2\frac{(\sigma,\nu_t)^2}{(\sigma^2,\nu_t)}}
\ee{pmatrix} \be{pmatrix}
\pi_t\sqrt{(\sigma^2,\nu_t)}&\rho\frac{(\sigma,\nu_t)}{\sqrt{(\sigma^2,\nu_t)}}\\[3mm]
0&\sqrt{1-\rho^2\frac{(\sigma,\nu_t)^2}{(\sigma^2,\nu_t)}}
\ee{pmatrix}\\
 =\be{pmatrix}
(\sigma^2,\nu_t)\pi_t^2&\rho(\sigma,\nu_t)\pi_t\\[3mm]
\rho(\sigma,\nu_t)\pi_t&1 \ee{pmatrix}
\end{align}
the generator of the process $(X_t,Y_t)$ can be given by the
function
\begin{multline}\label{ham}
\quad{\cal H}^{\pi,\mu,\sigma}(x,y,p,q)
\\=\frac{1}{2}\,\pi^2\sigma^2q_{11}+\rho\pi\sigma
q_{12}+\frac{1}{2}\,q_{22}+xr(y)p_1+\pi b(y)p_1+\pi \mu
p_1+\beta(y)p_2.\quad
\end{multline}
 For all $\nu\in {\cal P}(K), \pi\in R$
and $(x,y,p,q)\in R_+\times R\times R^2\times R^3$ we set
\begin{align}
 {\cal H}^{\pi,\nu}(x,y,p,q)&=({\cal
H}^{\pi,\mu,\sigma}(x,y,p,q),\nu)\\
\intertext{and}
{\cal H}(x,y,p,q)&=\max_{\pi\in \mathbb{R}}\min_{\nu\in {\cal
P}(K)}{\cal H}^{\pi,\nu}(x,y,p,q).
\end{align}

\begin{prop}\lbl{neu}
For each fixed  $(x,y,p,q)\in R_+\times R\times R^2\times R^3$,
with $q_{11}<0$ the function $(\pi,\nu)\to{\cal
H}^{\pi,\mu,\sigma}(x,y,p,q)$ admits a saddle point
$(\pi^*,\nu^*)$, i.e.
\begin{align}\label{ham30}
{\cal H}^{\pi^*,\nu^*}(x,y,p,q)=\max_{\pi\in
\mathbb{R}}\min_{\nu\in {\cal P}(K)}{\cal H}^{\pi,\nu}(x,y,p,q)
=\min_{\nu\in {\cal P}(K)}\max_{\pi\in \mathbb{R}}{\cal
H}^{\pi,\nu}(x,y,p,q).
\end{align}
Moreover
\begin{align}\lbl{more}
\max_{\pi\in \mathbb{R}}\min_{\nu\in {\cal P}(K)}{\cal
H}^{\pi,\nu}(x,y,p,q)=\max_{\pi\in \mathbb{R}}\min_{(\mu,\sg)\in
K}{\cal H}^{\pi,\mu,\sg}(x,y,p,q).
\end{align}
\end{prop}

{\it Proof}.
 By Theorem of Neumann at al.
(see Theorem lX.4.1 of \cite{var}) for each positive $n$ and fixed
point $(x,y,p,q)$ the function of measures $\ld\in {\cal
P}([-n,n]),\;\nu\in {\cal P}(K)$
$$(\ld,\nu)\to{\cal H}^{\ld,\nu}(x,y,p,q)=\int_{-n}^n\int_K{\cal
H}^{\pi,\mu,\sg}(x,y,p,q)\ld(d\pi)\nu(d\mu d\sg),$$ admits a
saddle point $(\ld_n^*,\nu_n^*)$, i.e.
\begin{align}{\cal H}^{\ld_n^*,\nu_n^*}(x,y,p,q)&=
\max_{\ld\in {\cal P}([-n,n])}\min_{\nu\in
{\cal P}(K)}{\cal H}^{\ld,\nu}(x,y,p,q)\notag\\
&=\min_{\nu\in {\cal P}(K)}\max_{\ld\in {\cal
P}([-n,n])}{\cal H}^{\ld,\nu}(x,y,p,q).
\end{align}
By concavity of ${\cal H}^{\pi,\nu}$ with respect to $\pi$ the
maximizer $\pi_n^*=\arg\min{\cal H}^{\pi,\nu}$ is unique and thus
$\ld_n^*=\delta_{\pi_n^*}$
 \footnote{$^)$ $\delta_{a}$ denotes a  measure with support in the point $a$}${}^)$.
Therefore we have
\begin{align}
\notag\pi_n^* =\begin{cases} \ds -n,\;\;&\ds {\rm
if}\;-\frac{b(y)p_1+(\mu,\nu_n^*)p_1\!+\!(\sg,\nu_n^*)\rho
q_{12}}{(\sg^2,\nu_n^*)q_{11}}<-n, \\
\ds -\frac{b(y)p_1\!+\!(\mu,\nu_n^*)p_1\!+\!(\sg,\nu_n^*)\rho
q_{12}}{(\sg^2,\nu_n^*)q_{11}},\;&\ds {\rm
if}\;-\frac{(b(y)p_1+\mu,\nu_n^*)p_1+(\sg,\nu_n^*)\rho
q_{12}}{(\sg^2,\nu_n^*)q_{11}}\in[-n,n]\\
n,\;\;&\ds {\rm
if}\;-\frac{(b(y)p_1+\mu,\nu_n^*)p_1+(\sg,\nu_n^*)\rho
q_{12}}{(\sg^2,\nu_n^*)q_{11}}>n
\end{cases}
\end{align}
and
\begin{align}\label{ham3}
{\cal H}^{\pi_n^*,\nu_n^*}& (x,y,p,q)=\max_{\pi\in
[-n,n]}\min_{\nu\in
{\cal P}(K)}{\cal H}^{\pi,\nu}(x,y,p,q)\notag \\
\notag&\equiv\frac{1}{2}q_{22}+\beta(y)p_2+xr(y)p_1\\
\notag &\quad + \max_{\pi\in [-n,n]}\min_{\nu\in {\cal
P}(K)}\left[\frac{1}{2}(\sigma^2,\nu)q_{11}\pi^2+(\sigma,\nu)\rho
q_{12}\pi+
(b(y)+(\mu,\nu))p_1\pi\right]\\
\notag&=\frac{1 }{2}q_{22}+\beta(y)p_2+xr(y)p_1\\
\notag &\quad +\min_{\nu\in {\cal P}(K)}\max_{\pi\in
[-n,n]}\left[\frac{1}{2}(\sigma^2,\nu)q_{11}\pi^2+(\sigma,\nu)\rho
q_{12}\pi+
(b(y)+(\mu,\nu))p_1\pi\right]\\
&\equiv\min_{\nu\in {\cal P}(K)}\max_{\pi\in [-n,n]}{\cal
H}^{\pi,\nu}(x,y,p,q).
\end{align}
By compactness of ${\cal P}(K)$ we can assume without loss of
generality that the sequence $\nu_n^*$ is convergent to some
$\nu^*$. Thus
$$
\pi_n^*\rightarrow
\pi^*\equiv-\frac{b(y)p_1+(\mu,\nu^*)p_1+(\sg,\nu^*)\rho
q_{12}}{(\sg^2,\nu^*)q_{11}}\;\;{\rm as}\;\;n\to\infty.
$$
It remains to use the equalities
\begin{multline}\lbl{h+}
\quad \max_{\pi\in \mathbb{R}}\min_{\nu\in {\cal P}(K)}{\cal
H}^{\pi,\nu}(x,y,p,q)\\
 =\lim_{n\to\infty}\max_{\pi\in
[-n,n]}\min_{\nu\in {\cal P}(K)}{\cal H}^{\pi,\nu}(x,y,p,q) ={\cal
H}^{\pi^*,\nu^*}(x,y,p,q)\quad
\end{multline}
and
\begin{multline}\lbl{h-}
\quad \min_{(\mu,\sg)\in K}\max_{\pi\in \mathbb{R}}{\cal
H}^{\pi,\mu,\sg}(x,y,p,q)\\
=\lim_{n\to\infty}\min_{(\mu,\sg)\in
K}\max_{\pi\in [-n,n]}{\cal H}^{\pi,\mu,\sg}(x,y,p,q)={\cal
H}^{\pi^*,\nu^*}(x,y,p,q) \quad
\end{multline}
to conclude that $(\pi^*,\nu^*)$ is saddle point of the problem.

On the other hand for each continuous function $f$ on $K$
$$\min_{\nu\in
{\cal P}(K)}(f,\nu)=\min_{(\mu,\sg)\in K}f(\mu,\sg),$$ since for
$\nu^*=\arg\min_\nu(f,\nu)$ we have ${\rm
supp}\nu^*\subseteq\{(\mu^*,\sg^*)|f(\mu^*,\sg^*)=\min
f(\mu,\sg)\}.$ \\
 Hence
\begin{align}
\max_{\pi\in [-n,n]}\min_{\nu\in
{\cal P}(K)}{\cal H}^{\pi,\nu}(x,y,p,q)
=\max_{\pi\in [-n,n]}\min_{(\mu,\sg)\in K}{\cal
H}^{\pi,\mu,\sg}(x,y,p,q).
\end{align}
This equality together to (\ref{h+}),(\ref{h-}) prove
(\ref{more}). \qed

Now we define the value functions
\begin{equation}
\begin{aligned}
v^-(t,x,y)=\max_{\pi\in \Pi_x}\min_{(\mu,\sigma)\in {\cal U}_K}
EU(X_T^{t,x,y})\\
v^+(t,x,y)=\min_{(\mu,\sigma)\in \bar{\cal
U}_K}\max_{\pi\in \Pi_x} EU(X_T^{t,x,y}).
\end{aligned}
\end{equation}
Since the Isaacs condition is satisfied (by Proposition \ref{neu})
there exists value of differential game  $v\equiv v^+=v^-$, which
would be solution of HJBI equation
\begin{gather}
\frac{\partial}{\partial t}v(t,x,y)+{\cal H}(t,y,v_x(t,x,y),v_y(t,x,y),v_{xx}(t,x,y),v_{xy}(t,x,y),v_{yy}(t,x,y))=0,\\
v(T,x,y)=U(x).
\end{gather}
It can be rewritten as
\begin{align}\lbl{hjb00}
& \frac{\partial}{\partial t}v(t,x,y)+\frac{1}{2}v_{yy}(t,x,y)+\beta(y)v_y(t,x,y)+xr(y)v_x(t,x,y)\notag \\
\notag &\quad  +\min_{\nu\in {\cal P}(K)}\max_{\pi\in
R}\big[\frac{1}{2}(\sigma^2,\nu)v_{xx}(t,x,y)\pi^2+(\sigma,\nu)\rho
v_{xy}(t,x,y)\pi\\
& \quad +
(b(y)+(\mu,\nu))v_{x}(t,x,y)\pi\big]=0,\\
& \lbl{hjb00ter} v(T,x,y)=U(x).
\end{align}

Simplifying the expression we have
\begin{align} %%%%%%%%%(2.27)-(2.29)
& \min_{\nu\in {\cal P}(K)}\max_{\pi\in
\mathbb{R}}\left[\frac{1}{2}(\sigma^2,\nu)q_{11}\pi^2+(\sigma,\nu)\rho
q_{12}\pi+b(y)p_1\pi+
(\mu,\nu)p_1\pi\right] \notag \\
=\, & \min_{\nu\in {\cal P}(K)}\left[\frac{((\sigma,\nu)\rho q_{12}+
(b(y)+(\mu,\nu))p_1)^2}{-2(\sigma^2,\nu)q_{11}}\right] \notag
\\
=\, & \be{cases} \ds -\frac{p_1^2}{2q_{11}}\min_{\nu\in {\cal
P}(K)}\left[\frac{((\sigma,\nu)\kappa+
b(y)+(\mu,\nu))^2}{(\sigma^2,\nu)}\right],\;\;&{\rm if}\;p_1\neq0\\[2mm]
\ds -\frac{\rho^2q_{12}^2}{2\sg_M},\;\;&{\rm if}\;p_1=0, \ee{cases}
\end{align}
where we suppose that $q_{11}<0$ and use the notation
$\kappa=\frac{\rho q_{12}}{p_1}$.

For the sake of simplicity  we assume in addition

A3) $b(y)+\mu_-\ge0,\;\text{\rm for all}\;y\in \mathbb{R}.$

 \be{prop} There exists $\nu^*\in{\cal P}(K)$ of the form
 $\nu^*=\alpha\delta_{\mu_\pm,\sg_-}+(1-\alpha)\delta_{\mu_\pm,\sg_+}$, $0\le\alpha\le1$,
such that

\beq \min_{\nu\in {\cal
P}(K)}\left[\frac{((b(y)+\mu,\nu)+\kap(\sigma,\nu))^2}{(\sigma^2,\nu)}\right]=\frac{((b(y)+\mu,\nu^*)+\kap(\sigma,\nu^*))^2}{(\sigma^2,\nu^*)}
\eeq and
\begin{equation}\lbl{nu}
((\mu,\nu^*),(\sigma,\nu^*))=\be{cases}
(\mu_+,\frac{\mu_+}{\kap}+\frac{\sg_-\sg_+}{\sg_M}),\;\;&\ds {\rm if}\;\kappa\in\left(-\infty,\frac{\mu_+\sg_M}{\sg_M\sg_--\sg_+\sg_-}\right]\\[3mm]
(\mu_+,\sg_-),\;&\ds {\rm if}\;\;\kappa\in\left(\frac{\mu_+\sg_M}{\sg_M\sg_--\sg_+\sg_-},-\frac{\mu_+}{\sg_-}\right]\\[3mm]
(\kap,-1)\;{\rm constant},\;\;&\ds {\rm if}\;\kappa\in\left(-\frac{\mu_+}{\sg_-},-\frac{\mu_-}{\sg_+}\right]\\[3mm]
(\mu_-,\sg_+),\;\;&\ds {\rm if}\;\kap\in\left(-\frac{\mu_-}{\sg_+},\frac{\mu_-\sg_M}{\sg_M\sg_+-\sg_+\sg_-}\right]\\[3mm]
(\mu_-,\frac{\mu_-}{\kap}+\frac{\sg_-\sg_+}{\sg_M}),\;\;&\ds {\rm
if}\;\kap\in\left(\frac{\mu_-\sg_M}{\sg_M\sg_+-\sg_+\sg_-},\infty\right)
\ee{cases},
\end{equation}
\begin{multline}
\frac{((b(y)+\mu,\nu^*)+\kap(\sigma,\nu^*))^2}{(\sigma^2,\nu^*)}\\
=\be{cases}
\ds \frac{\kap(2(b(y)+\mu_+)\sg_M+\kappa\sg_-\sg_+)}{\sg_M^2},\;\;&\ds {\rm if}\;\kappa\in\left(-\infty,\frac{\mu_+\sg_M}{\sg_M\sg_--\sg_+\sg_-}\right]\\[3mm]
\ds \frac{(b(y)+\mu_++\kappa\sg_-)^2}{\sg_-^2},\;&\ds {\rm if}\;\;\kappa\in\left(\frac{\mu_+\sg_M}{\sg_M\sg_--\sg_+\sg_-},-\frac{\mu_+}{\sg_-}\right]\\[3mm]
0,\;\;&\ds {\rm if}\;\kappa\in\left(-\frac{\mu_+}{\sg_-},-\frac{\mu_-}{\sg_+}\right]\\[3mm]
\ds \frac{(b(y)+\mu_-+\kappa\sg_+)^2}{\sg_+^2},\;\;&\ds {\rm if}\;\kap\in\left(-\frac{\mu_-}{\sg_+},\frac{\mu_-\sg_M}{\sg_M\sg_+-\sg_+\sg_-}\right]\\[3mm]
\ds \frac{\kap(2(b(y)+\mu_-)\sg_M+\kappa\sg_-\sg_+)}{\sg_M^2},\;\;&\ds {\rm
if}\;\kap\in\left(\frac{\mu_-\sg_M}{\sg_M\sg_+-\sg_+\sg_-},\infty\right)
\ee{cases}.
\end{multline}
\ee{prop}

The proof is given in Appendix.

\be{cor}\lbl{crr}
\begin{align}
\min_{\nu\in {\cal
P}(K)}\left[\frac{((b(y)+\mu,\nu)p_1+(\sigma,\nu)\rho q_{12}
)^2}{2(\sigma^2,\nu)q_{11}}\right]= \min_{(\mu,\sg)\in
K}\left[\frac{(b(y)p_1+\mu p_1+\sigma\rho q_{12}
)^2}{2(2\sg_M\sg-\sg_-\sg_+)q_{11}}\right] \notag \\[3mm]
= \be{cases}
\ds \frac{\rho q_{12}(2p_1(b(y)+\mu_+)\sg_M+\rho
q_{12}\sg_-\sg_+)}{2q_{11}\sg_M^2},\;\;&\ds {\rm if}\;
\frac{\rho q_{12}}{p_1}\in\left(-\infty,\frac{\mu_+\sg_M}{\sg_M\sg_--\sg_+\sg_-}\right]\\[3mm]
\ds \frac{(p_1(b(y)+\mu_+)+\rho
q_{12}\sg_-)^2}{2q_{11}\sg_-^2},\;\;&\ds {\rm if}\;
\frac{\rho q_{12}}{p_1}\in\left(\frac{\mu_+\sg_M}{\sg_M\sg_--\sg_+\sg_-},-\frac{\mu_+}{\sg_-}\right]\\[3mm]
0,\;\;&\ds {\rm if}\;\frac{\rho q_{12}}{p_1}\in\left(-\frac{\mu_+}{\sg_-},-\frac{\mu_-}{\sg_+}\right]\\[3mm]
\ds \frac{(p_1(b(y)+\mu_-)+\rho
q_{12}\sg_+)^2}{2q_{11}\sg_+^2},\;\;&\ds {\rm if}\;
\frac{\rho q_{12}}{p_1}\in\left(-\frac{\mu_-}{\sg_+},\frac{\mu_-\sg_M}{\sg_M\sg_+-\sg_+\sg_-}\right]\\[3mm]
\ds \frac{\rho q_{12}(2p_1(b(y)+\mu_-)\sg_M+\rho
q_{12}\sg_-\sg_+)}{2q_{11}\sg_M^2},\;\;&\ds {\rm if}\;
\frac{\rho q_{12}}{p_1}\in\left(\frac{\mu_-\sg_M}{\sg_M\sg_+-\sg_+\sg_-},\infty\right)\\[3mm]
\frac{\rho^2q_{12}^2}{2\sg_M},\;\;&{\rm if}\;p_1=0 \ee{cases}.
\end{align} \ee{cor}

{\it Proof}. It is sufficient to verify that for
$\nu^*=\alpha\delta_{\mu_\pm,\sg_-}+(1-\alpha)\delta_{\mu_\pm,\sg_+},\;0\le\alpha\le1$
we get  $(\sg^2,\nu^*)=2\sg_M(\sg,\nu^*)-\sg_-\sg_+$.

From this Corollary we obtain that the HJBI equation has the
form
\begin{align}\lbl{hjb0}
& \frac{\partial}{\partial t}v(t,x,y)+\frac{1}{2}v_{yy}(t,x,y)+\beta(y)v_y(t,x,y)+xr(y)v_x(t,x,y)\notag \\
& \quad -\min_{(\mu,\sg)\in K}\frac{(b(y)v_x(t,x,y)+\mu v_x(t,x,y)+\rho\sg v_{xy}(t,x,y))^2}{2(2\sg_M\sg-\sg_-\sg_+)v_{xx}(t,x,y)}=0,\\
& \lbl{hjb0ter} v(T,x,y)=U(x).
\end{align}

Following to the Theorem 6 of {\cite{nis} we can prove

\be{thr}[Verification Theorem] \lbl{veri}
Let $v(t,x,y)$ be a
classical solution of $(\ref{hjb00})$, $(\ref{hjb00ter})$ with
$v_{xx}<0$. Then there exists $\nu^*$ defined by $(\ref{nu})$ with
$\kap=\rho\frac{v_{xy}}{v_x}$ and the optimal strategy is given by
\beq\lbl{pi*}
\pi^*(t,x,y)=-\frac{(b(y)+\mu,\nu^*(t,x,y))v_x(t,x,y)+(\sg,\nu^*(t,x,y))\rho
v_{xy}(t,x,y)}{(2\sg_M(\sg,\nu^*(t,x,y)))-\sg_-\sg_+)v_{xx}(t,x,y)},
\eeq
where
\begin{multline}\lbl{nu*}
((\mu,\nu^*(t,x,y)),(\sigma,\nu^*(t,x,y)))\\
=\be{cases}
\ds \left(\mu_+,\frac{\mu_+v_x(t,x,y)}{\rho
v_{xy}(t,x,y)}+\frac{\sg_-\sg_+}{\sg_M}\right),\;\;&\ds {\rm if}\;
\frac{\rho v_{xy}(t,x,y)}{v_x(t,x,y)}\in\left(-\infty,\frac{\mu_+\sg_M}{\sg_M\sg_--\sg_+\sg_-}\right]\\[3mm]
(\mu_+,\sg_-),\;&\ds {\rm if}\;\frac{\rho v_{xy}(t,x,y)}{v_x(t,x,y)}\in\left(\frac{\mu_+\sg_M}{\sg_M\sg_--\sg_+\sg_-},-\frac{\mu_+}{\sg_-}\right]\\[3mm]
\ds \left(\frac{\rho v_{xy}(t,x,y)}{v_x(t,x,y)},-1\right){\rm constant},\;\;&\ds {\rm if}\;\frac{\rho v_{xy}(t,x,y)}{v_x(t,x,y)}\in\left(-\frac{\mu_+}{\sg_-},-\frac{\mu_-}{\sg_+}\right]\\[3mm]
(\mu_-,\sg_+),\;\;&\ds {\rm if}\;\frac{\rho v_{xy}(t,x,y)}{v_x(t,x,y)}\in\left(-\frac{\mu_-}{\sg_+},\frac{\mu_-\sg_M}{\sg_M\sg_+-\sg_+\sg_-}\right]\\[3mm]
\ds \left(\mu_-,\frac{\mu_+v_x(t,x,y)}{\rho
v_{xy}(t,x,y)}+\frac{\sg_-\sg_+}{\sg_M}\right),\;\;&\ds {\rm if}\;\frac{\rho
v_{xy}(t,x,y)}{v_x(t,x,y)}\in\left(\frac{\mu_-\sg_M}{\sg_M\sg_+-\sg_+\sg_-},\infty\right)
\ee{cases}.
\end{multline}
\ee{thr}

\

\section{The power utility case}

\

We now consider the robust utility maximization problem with power
utility $U(x)=\frac{1}{q}x^q,$ with $q<1,\;q\neq0$. Hence we
obtain the equation
\begin{align}\lbl{hjb1}
\frac{\partial}{\partial t}v(t,x,y)+\frac{1}{2}v_{yy}(t,x,y)+\beta(y)v_y(t,x,y)+xr(y)v_x(t,x,y)\notag \\ -\min_{(\mu,\sg)\in K}\frac{((b(y)+\mu) v_x(t,x,y)+\rho\sg v_{xy}(t,x,y))^2}{2(2\sg_M\sg-\sg_-\sg_+)v_{xx}(t,x,y)}=0,\\
v(T,x,y)=\frac{1}{q}x^q.
\end{align}
The solution of this equation is of the form
$v(t,x,y)=\frac{1}{q}x^qe^{u(t,y)}$, where $u$ satisfies
\begin{align}\lbl{hjb2}
\frac{\partial}{\partial t}u(t,y)+\frac{1}{2}u_{yy}(t,y)+\beta(y)u_y(t,y)+\frac{1}{2}u_y^2(t,y)+qr(y)\notag \\
-\frac{1}{2(q-1)}\min_{(\mu,\sg)\in K}\frac{(b(y)+\mu +\rho\sg u_y(t,y))^2}{2\sg_M\sg-\sg_-\sg_+}=0,\\
\lbl{hjb2ter}u(T,y)=0.
\end{align}
It is evident that $v_{xx}(t,x,y)=(q-1)x^{q-2}e^{u(t,y)}<0$.

The equations (\ref{nu*}) take the form
\begin{multline}\lbl{nu**}
((\mu,\nu^*(t,y)),(\sigma,\nu^*(t,y)))\\
=\be{cases}
\ds \left(\mu_+,\frac{\mu_+}{\rho}u_y(t,y)+\frac{\sg_-\sg_+}{\sg_M}\right),\;\;&\ds {\rm
if}\;
\rho u_y(t,y)\in\left(-\infty,\frac{\mu_+\sg_M}{\sg_M\sg_--\sg_+\sg_-}\right]\\[3mm]
(\mu_+,\sg_-),\;&\ds {\rm if}\;\rho u_y(t,y)\in \left(\frac{\mu_+\sg_M}{\sg_M\sg_--\sg_+\sg_-},-\frac{\mu_+}{\sg_-}\right]\\[3mm]
(\rho u_y(t,y),-1){\rm constant},\;\;&\ds {\rm if}\;\rho u_y(t,y)\in \left(-\frac{\mu_+}{\sg_-},-\frac{\mu_-}{\sg_+}\right]\\[3mm]
(\mu_-,\sg_+),\;\;&\ds {\rm if}\;\rho u_y(t,y)\in \left(-\frac{\mu_-}{\sg_+},\frac{\mu_-\sg_M}{\sg_M\sg_+-\sg_+\sg_-}\right]\\[3mm]
\ds (\mu_-,\frac{\mu_+}{\rho}u_y(t,y)+\frac{\sg_-\sg_+}{\sg_M}),\;\;&\ds {\rm
if}\;\rho
u_y(t,y)\in\left(\frac{\mu_-\sg_M}{\sg_M\sg_+-\sg_+\sg_-},\infty\right)
\ee{cases}.
\end{multline}

\be{rem} By corollary \ref{crr}  and (\ref{nu*}) the equation
(\ref{hjb2}) can be written as
\begin{align}\lbl{hjb11}
& \frac{\partial}{\partial t}u(t,y)+\frac{1}{2}u_{yy}(t,y)+\beta(y)u_y(t,y)+\frac{1}{2}u_y^2(t,y)+qr(y)\notag \\
\notag &  -\frac{\rho
u_y(t,y)}{2(q-1)\sg_M^2}{(2(b(y)+\mu_+)\sg_M+\sg_-\sg_+\rho
u_y(t,y))} \chi\left(\rho u_y(t,y)\le
\frac{\mu_+\sg_M}{\sg_M\sg_--\sg_+\sg_-}
\right)\\
\notag&  -\frac{1}{2(q-1)\sg_-^2}{(b(y)+\mu_++\rho\sg_-u_y(t,y))^2}\chi\left(\frac{\mu_+\sg_M}{\sg_M\sg_--\sg_+\sg_-}
<\rho u_y(t,y)
\le-\frac{\mu_+}{\sg_-}\right)\\
\notag & \; -\frac{1}{2(q-1)\sg_+^2}{(b(y)+\mu_-+\rho\sg_+u_y(t,y))^2}\chi\left(-\frac{\mu_-}{\sg_+}
<\rho u_y(t,y)\le \frac{\mu_-\sg_M}{\sg_M\sg_+-\sg_+\sg_-}
\right)\\
& -\frac{\rho
u_y(t,y)}{2(q\!-\!1)\sg_M^2}{(2(b(y)+\mu_-)\sg_M\!+\!\sg_-\sg_+\rho
u_y(t,y))} \chi\left(\rho
u_y(t,y)\!>\!\frac{\mu_-\sg_M}{\sg_M\sg_+\!-\!\sg_+\sg_-}
\right)=0,\\
& \hskip+12.5cm u(T,y)=0,
\end{align}
where $\chi$ is the characteristic function.
\ee{rem}

 \be{thr} Under conditions A1)-A3) the problem
$(\ref{hjb2})$,\,$(\ref{hjb2ter})$ admits a classical solution
with bounded $u_y(t,y)$ and a saddle point
$(\nu^*(t,y),\pi^*(t,x,y))$ of the problem
$(\ref{mima})$,\,$(\ref{cap2})$  is defined by $(\ref{nu**})$ and
by
 \beq
\pi^*(t,x,y)=\frac{x}{1-q}\left(\frac{b(y)+(\mu,\nu^*(t,y))}{(\sg^2,\nu^*(t,y))}+\rho\frac{(\sg,\nu^*(t,y))}{(\sg^2,\nu^*(t,y))}
u_{y}(t,y)\right). \eeq \ee{thr}

{\it Proof}. By condition A2) there exists $N\ge0$ such that
$b'(y)=0$, if $|y|>N$. Thus $b(y)=b^+$, if $y\ge N$ and
$b(y)=b^-$, if $y\le -N$ for some constants $b^+,b^-$. The
solution of (\ref{hjb2}) on the intervals $(-\infty,-N]$ and
$[N,\infty)$ are
$u^-(t)=-\frac{1}{2(q-1)}\frac{(b^-+\mu_-)^2}{\sg_+^2}(T-t)$ and
$u^+(t)=-\frac{1}{2(q-1)}\frac{(b^++\mu_-)^2}{\sg_+^2}(T-t)$
respectively. Now we consider the Cauchy-Dirichlet problem on the
bounded domain $(0,T)\times(-N,N)$
\begin{align}\lbl{hjb22}
\frac{\partial}{\partial t}u(t,y)+\frac{1}{2}u_{yy}(t,y)+\beta(y)u_y(t,y)+\frac{1}{2}u_y^2(t,y)\notag \\
-\frac{1}{2(q-1)}\min_{(\mu,\sg)\in K}\frac{(b(y)+\mu +\rho\sg u_y(t,y))^2}{2\sg_M\sg-\sg_-\sg_+}=0,\\
\lbl{hjb22ter}u(T,y)=0,\;u(t,\pm N)=u^\pm(t).
\end{align}
 Suppose
$$a_1(t,y,u,p)=\frac{1}{2}p,
$$
$$
a(t,y,u,p)=\beta(y)p+\frac{1}{2}p-\frac{1}{2(q-1)}\min_{(\mu,\sg)\in
K}\frac{(b(y)+\mu +\rho\sg p)^2}{2\sg_M\sg-\sg_-\sg_+}.
$$
 It is
easy to see that $a$ is  Lipschitz function on the each ball of
its domain, $a(t,y,u,0)$ is bounded below and all conditions of
Theorem 6.2 chapt.V of (\cite{L-S-U}) are satisfied. Therefore
there exists a classical solution of (\ref{hjb2}),(\ref{hjb2ter})
with bounded $u_y(t,y)$ (the existence of classical solution
follows also from Example 3.6 of \cite{hen} if we consider mixed
problem with boundary conditions $u(T,y)=0,\;u_y(t,\pm N)+u(t,\pm
N)=u^\pm(t)$).

Now we can use the Theorem \ref{veri}. From (\ref{pi*}) follows
that the strategy is of the form

\begin{align} \notag
\pi^*(t,x,y)=-\frac{1}{q-1}\frac{b(y)+(\mu,\nu^*(t,y))+(\sg,\nu^*(t,y))\rho
u_{y}(t,y)}{2(\sg,\nu^*(t,y))\sg_M-\sg_-\sg_+}x\\
=\frac{1}{1-q}\frac{b(y)+(\mu,\nu^*(t,y))+(\sg,\nu^*(t,y))\rho
u_{y}(t,y)}{(\sg^2,\nu^*(t,y))}x,\end{align}

where $\nu^*(t,y)$ is defined by (\ref{nu**}).
 \be{cor}
If $b=0$ then
$$u(t,y)=-\frac{1}{2(q-1)}(T-t)\min_{(\mu,\sg)\in
K}\frac{\mu^2}{2\sg_M\sg-\sg_-\sg_+}=-\frac{1}{2(q-1)}(T-t)\frac{\mu_-^2}{\sg_+^2}$$
is a solution of $(\ref{hjb2})$ and a saddle point of the maximin
problem can be given explicitly
$$(\mu_t^*,\sg_t^*)=(\mu_-,\sg_+),\;\;\pi^*(t,x,y)=-\frac{\mu_-}{2(q-1)\sg_+^2}x.$$
\ee{cor}

\be{rem}
 When $\sg_-=\sg_+=\sg_M$ we obtain
\begin{align}\lbl{hjb3}
& \frac{\partial}{\partial t}u(t,y)+\frac{1}{2}u_{yy}(t,y)+\beta(y)u_y(t,y)+\frac{1}{2}u_y^2(t,y)\notag \\
& \qquad \notag-\frac{1}{2(q-1)\sg_M^2}\min_{\mu_-\le\mu\le\mu_+}(b(y)+\mu
+\rho\sg_M u_y(t,y))^2\\
& \quad \notag \equiv\frac{\partial}{\partial t}u(t,y)+
\frac{1}{2}u_{yy}(t,y)+(2\rho\sg_Mb(y)+\beta(y))u_y(t,y)+\frac{1}{2}\left(1-\frac{\rho^2\sg_M}{q-1}\right)u_y^2(t,y)\\
& \qquad -\frac{1}{2(q-1)\sg_M^2}\min_{\mu_-\le\mu\le\mu_+}((b(y)+\mu)^2
+2\mu\rho\sg_M u_y(t,y))=0,\\
& \hskip+9.1cm u(T,y)=0.
\end{align}
 The existence of classical solution of
such type equation has been obtained by D. Hern\'an\-dez-Hern\'andez
and A. Schied in \cite{herher}. \ee{rem}

\be{rem} Instead of PDE (\ref{hjb2}) we can use BSDE with
quadratic growth
\begin{align}\lbl{hjbbs}
dV_t& =-\bigg(\beta(w_t)Z_t+\frac{1}{2}Z_t^2+qr(w_t)\notag \\
& \quad -\frac{1}{2(q-1)}\min_{(\mu,\sg)\in K}\frac{(b(w_t)+\mu +\rho\sg Z_t)^2}{2\sg_M\sg-\sg_-\sg_+}\bigg)dt+Z_tdw_t+Z_t^\bot dw_t^\bot,\\
V_T& =0.
\end{align}
which solvability follows from the results of \cite{Kob},
\cite{Tev}. The strategy now is a solution of forward SDE
 \beq
\pi_t^*=\frac{1}{1-q}\left(\frac{b(w_t)+(\mu,\nu_t^*(Z))}{(\sg^2,\nu_t^*(Z))}+\rho\frac{(\sg,\nu_t^*(Z))}{(\sg^2,\nu_t^*(Z))}
Z_t\right)X_t(\pi^*). \eeq
 \ee{rem}

\

\appendix
\section{Appendix}

Each measure $\nu$ may be realized as a distribution of the pair
of random variables $(\xi,\eta)$ with the value in $D$.
Simplifying the notation we $b(y)+\mu$ denote again by $\mu$. Our
aim is to characterize the dependence of the minimizer of the
problem \beq \min_{\nu\in {\cal
P}(K)}\left[\frac{((\mu,\nu)+\kap(\sigma,\nu))^2}{(\sigma^2,\nu)}\right]=\min_{(\xi,\eta)\in
K}\left[\frac{(E\xi+\kap E\eta)^2}{E\eta^2}\right] \eeq on the
parameter $\kap\in \mathbb{R}$.

\be{prop} The pair
$$(\xi^*,\eta^*)=\arg\min_{(\xi,\eta)\in K}\left[\frac{(E\xi+\kap
E\eta)^2}{E\eta^2}\right]$$ is such that $\xi^*$ is number,
$\eta^*$ is Bernoulli random variables with value
$\{\sg_-,\sg_+\}$ and their expectations are given as
\begin{equation}
(\xi^*,E\eta^*)=\be{cases}
\ds \left(\mu_+,\frac{\mu_+}{\kap}+\frac{\sg_-\sg_+}{\sg_M}\right),\;\;&\ds {\rm if}\;
\kap\in\left(-\infty,\frac{\mu_+\sg_M}{\sg_M\sg_--\sg_+\sg_-}\right]\\[3mm]
(\mu_+,\sg_-),\;&\ds {\rm if}\;\;\kap\in\left(\frac{\mu_+\sg_M}{\sg_M\sg_--\sg_+\sg_-},-\frac{\mu_+}{\sg_-}\right]\\[3mm]
(\kap,-1){\rm constant},\;\;&\ds {\rm if}\;\kap\in\left(-\frac{\mu_+}{\sg_-},-\frac{\mu_-}{\sg_+}\right]\\[3mm]
(\mu_-,\sg_+),\;\;&\ds {\rm if}\;\kap\in\left(-\frac{\mu_-}{\sg_+},\frac{\mu_-\sg_M}{\sg_M\sg_+-\sg_+\sg_-}\right]\\[3mm]
\ds \left(\mu_-,\frac{\mu_-}{\kap}+\frac{\sg_-\sg_+}{\sg_M}\right),\;\;&\ds {\rm
if}\;\kap\in\left(\frac{\mu_-\sg_M}{\sg_M\sg_+-\sg_+\sg_-},\infty\right)
\ee{cases}.
\end{equation}
Moreover
\begin{equation}
\frac{(\xi^*+\kap E\eta^*)^2}{E\eta^{*2}}=\be{cases}
\ds \frac{\kap(2\mu_+\sg_M+\kap\sg_-\sg_+)}{\sg_M^2},\;\;&\ds {\rm if}\;\kap\in\left(-\infty,\frac{\mu_+\sg_M}{\sg_M\sg_--\sg_+\sg_-}\right]\\[3mm]
\ds \frac{(\mu_++\kap\sg_-)^2}{\sg_-^2},\;&\ds {\rm if}\;\kap\in\left(\frac{\mu_+\sg_M}{\sg_M\sg_--\sg_+\sg_-},-\frac{\mu_+}{\sg_-}\right]\\[3mm]
0,\;\;&\ds {\rm if}\;\kap\in\left(-\frac{\mu_+}{\sg_-},-\frac{\mu_-}{\sg_+}\right]\\[3mm]
\ds \frac{(\mu_-+\kap\sg_+)^2}{\sg_+^2},\;\;&\ds {\rm if}\;\kap\in\left(-\frac{\mu_-}{\sg_+},\frac{\mu_-\sg_M}{\sg_M\sg_+-\sg_+\sg_-}\right]\\[3mm]
\ds \frac{\kap(2\mu_-\sg_M+\kap\sg_-\sg_+)}{\sg_M^2},\;\;&\ds {\rm
if}\;\kap\in\left(\frac{\mu_-\sg_M}{\sg_M\sg_+-\sg_+\sg_-},\infty\right)
\ee{cases}.
\end{equation}
\ee{prop} {\it Proof.} Let
$(\mu_++\kap\sg_-)(\mu_-+\kap\sg_+)\le0$. Then by continuity of
function $\mu+\kap\sg, (\mu,\sg)\in K$, there exists
$(\hat\mu,\hat\sg)$ such that $\hat\mu+\kap\hat\sg=0$. Thus
$(\hat\mu,\hat\sg)\propto(\kap,-1)$ and $\left[\frac{(E\xi^*+\kap
E\eta^*)^2}{E\eta^{*2}}\right]=0$. If
$(\mu_++\kap\sg_-)(\mu_-+\kap\sg_+)>0$ then either
$\kap>\frac{\mu_-}{\sg_+}$ and $\xi^*=\mu_-$ or
$\kap<-\frac{\mu_+}{\sg_-}$ and $\xi^*=\mu_+$. Thus it is
sufficient to study the minimization problem

$$\min_{\eta\in [\sg_-,\sg_+]}\left[\frac{(\mu_a+\kap
E\eta)^2}{E\eta^2}\right]\;\;\text{\rm for}\;\; a=+,-.
$$

We will show  that $\eta^*$ is of the form
$\eta^*=\sg_-\chi_B+\sg_+\chi_{B^c}$ for some event $B$. Indeed,
if $E\eta^*=y$ then $E\eta^{*2}=2\sg_My-\sg_-\sg_+$ and $\eta^*$
is maximizer of the problem
$$\max_{\eta,E\eta=y}E\eta^2,$$
since for any $\eta$, with $E\eta=y$ we get
$$
E\eta^2=E(\eta-\sg_M)^2+2\sg_My-\sg_M^2
$$
$$
\le\left(\frac{\sg_+-\sg_-}{2}\right)^2+2\sg_My-\sg_M^2
$$
$$
=2\sg_My-\sg_-\sg_+=E\eta^{*2}.
$$
 Hence
$$\min_{\eta\in
[\sg_-,\sg_+]}\left[\frac{(\mu_a+\kap
E\eta)^2}{E\eta^2}\right]=\min_{\sg_-\le y\le\sg_+}\psi_a(y),$$
where $\psi_a(y)=\frac{(\mu_a+\kap y)^2}{2\sg_My-\sg_-\sg_+}$.
Since
$$\psi_a'(y)=\frac{\kap^2}{2\sg_M}-\frac{\kap^2}{2\sg_M}\frac{(2\sg_M\frac{\mu_a}{\kap}+\sg_-\sg_+)^2}{(2\sg_My-\sg_-\sg_+)^2}$$
the equation $\psi_a'(y)=0$ has two roots;
$$y_1^a=-\frac{\mu_a}{\kap},\;y_2^a=\frac{\mu_a}{\kap}+\frac{\sg_-\sg_+}{\sg_M}.$$
If $y_1^a=-\frac{\mu_a}{\kap}\in[\sg_-,\sg_+]$ then
$y_2^a=\frac{\mu_a}{\kap}+\frac{\sg_-\sg_+}{\sg_M}\in[-\sg_++\frac{\sg_-\sg_+}{\sg_M},-\sg_-+\frac{\sg_-\sg_+}{\sg_M}]$
and vise versa. Moreover
$[\sg_-,\sg_+]\cap[-\sg_++\frac{\sg_-\sg_+}{\sg_M},-\sg_-+\frac{\sg_-\sg_+}{\sg_M}]=\emptyset.$
Since $\lim_{y\to\pm\infty}\psi_a(y)=\pm\infty$ then the least
root is the maximizer and highest root is the minimizer. The case
of $y_1^a\in[\sg_-,\sg_+]$ is equivalent to
$$\kap\in[-\frac{\sg_+}{\mu_a},-\frac{\sg_-}{\mu_a}]$$ and gives
$\min\psi_a(y)=\psi_a(y_1^a)=0$. From the relation
$y_2^a\in[\sg_-,\sg_+]$ follows $-\sg_++\frac{\sg_-\sg_+}{\sg_M}
\le-\frac{\mu_a}{\kap}\le-\sg_+-\frac{\sg_-\sg_+}{\sg_M}$ which
equivalent to
$$\kap\in(-\infty,\frac{\mu_a}{\sg_--\frac{\sg_-\sg_+}{\sg_M}}]\cup[\frac{\mu_a}{\sg_+-\frac{\sg_-\sg_+}{\sg_M}},\infty).$$
In this case $\min_{\sg_-\le
y\le\sg_+}\psi_a(y)=\psi_a(y_2^a)=\kap\frac{2\mu_a+\kap\sg_-\sg_+}{\sg_M^2}$.

Now we will consider step by step the all possibilities of
displacement of $\kap$ in the intervals formulated in Proposition.

1)
$\kap\in(-\infty,\frac{\mu_a}{\sg_--\frac{\sg_-\sg_+}{\sg_M}}]$.
Since
$\frac{\mu_a}{\sg_--\frac{\sg_-\sg_+}{\sg_M}}\le-\frac{\mu_+}{\sg_-}$
then $\kap\in(-\infty,-\frac{\mu_+}{\sg_-}]$ and $\xi^*=\mu_+$.
Moreover
$\min\psi_+(y)=\psi_+(y_2^+)=\kap\frac{2\mu_++\kap\sg_-\sg_+}{\sg_M^2}$.

2)
$\kap\in(\frac{\mu_+}{\sg_--\frac{\sg_-\sg_+}{\sg_M}},-\frac{\mu_+}{\sg_-}]$.
From $\kap\le-\frac{\mu_+}{\sg_-}$ follows that
$y_1^+=-\frac{\mu_+}{\kap}<\sg_-$ and from
$\kap>\frac{\mu_+}{\sg_--\frac{\sg_-\sg_+}{\sg_M}}$ follows
$y_2^+=\frac{\mu_+}{\kap}+\frac{\sg_-\sg_+}{\sg_M}<\sg_-$. Hence
$\psi_+(y)$ is increasing on $[\sg_-,\sg_+]$ and
$\arg\min_{\sg_-\le y\le\sg_+}\psi_+(y)=\sg_-.$

3) $\kap\in(-\frac{\mu_+}{\sg_-},-\frac{\mu_-}{\sg_+}]$. Then
$y_1^+=-\frac{\mu_+}{\kap}\in[\sg_-,\sg_+]$ and $\min\psi_+(y)=0$.

4)
$\kap\in(-\frac{\mu_-}{\sg_+},\frac{\mu_-}{\sg_+-\frac{\sg_-\sg_+}{\sg_M}}]$.
Then $\frac{\mu_-}{\kap}>\sg_+-\frac{\sg_-\sg_+}{\sg_M}$ and
$y_1^-=-\frac{\mu_-}{\kap}<-\sg_++\frac{\sg_-\sg_+}{\sg_M}<\sg_-,
\;y_2^-=\frac{\mu_-}{\kap}+\frac{\sg_-\sg_+}{\sg_M}>\sg_+.$ Hence
$\psi_-(y)$ is decreasing on $[\sg_-,\sg_+]$ and
$\arg\min\psi_+(y)=\sg_+.$

5) $\kap\in(\frac{\mu_-}{\sg_+-\frac{\sg_-\sg_+}{\sg_M}},\infty]$.
Then $\kap>\frac{\mu_-}{\sg_+}$ and $\xi^*=\mu_-.$ On the other
hand from $\frac{\mu_-}{\kap}<\sg_+-\frac{\sg_-\sg_+}{\sg_M}$
follows $y_2^-\in[\sg_-,\sg_+].$ Hence $\min_{\sg_-\le
y\le\sg_+}\psi_-(y)=\psi_-(y_2^-).$

\

\end{document}